# Challenges in Maritime Education with Emerging Technologies: An Overview Through Maritime Firefighting

Hasan Ahmad [1*], Pedro Merino-Laso [1,2]

[1] French Maritime Academy (ENSM)
[2] French Naval Academy Research Institute, IRENav, EA 3634, Brest, France
* Corresponding author: e-mail: ahmad.hasan@supmaritime.fr

**Contribution category**:

- Funded project

**Introduction:** The rapid emergence of Artificial Intelligence (AI), Internet of Things (IoT), Unmanned Aerial Vehicles (UAVs), and interoperable maritime data systems is transforming maritime operations and emergency management (Boubendir et al., 2021; Wu et al., 2022). Maritime fire incidents involving dangerous goods (DG) and lithium-ion batteries expose limitations in conventional firefighting training and preparedness, creating new challenges for Maritime Education and Training (MET) under the evolving STCW framework particularly for simulator-based and digital competency development (Kim et al., 2021). **Aims:** This paper aims to identify the educational and competency challenges arising from emerging technologies in maritime firefighting, using the OVERHEAT Digital Solution (DS) as a practical case. It provides an initial identification of the knowledge, operational and technology-related competencies for future simulator-supported MET. **Methods:** OVERHEAT DS integrates smart sensors, UAV, intelligent data management, and ship-shore interoperability using Service-Oriented Architecture (SOA) and S-100 standards (Ahmad & Merino-Laso, 2025). Validation is progressively conducted through laboratory experiments, simulator-based demonstrations, and hybrid environments. ENSM's MARINS FMBS and open-source SOMOS simulator support the evaluation of interoperability, collaborative emergency-response workflows, and shared situational awareness (SSA) under simulated fire scenarios. **Results:** Initial experiments demonstrated early fire detection, while UAV provided real-time thermal and optical monitoring. Simulator-supported environments enabled testing of ECDIS/VTS interoperability, collaborative decision-making, and ship-shore coordination through integrated Common Operational Pictures (COPs) (Ahmad et al., 2025). The project also identified several educational challenges, related to S-100 integration, real-time data management and new digital competencies. **Conclusion:** Emerging technologies are reshaping maritime operations and creating new requirements for MET. The findings show that integrating AI, IoT, UAVs, and interoperable maritime data systems can enhance maritime firefighting and operational preparedness while highlighting the importance of adapting future MET and STCW-related competencies for digitalized and collaborative operations.

# 1. INTRODUCTION

Maritime safety is strongly influenced by the interaction between people, technology and operational environments (Mallam et al., 2019). Human factors (HF) remain central because competencies developed through education and training influence how seafarers perceive, interpret and respond to operational situations. Türkistanli (2023) highlights the relationship between MET, human error and safety, identifying situational awareness (SA), teamwork, communication, leadership and decision-making as important human-factor themes. These competencies are particularly relevant to safety-critical operations such as maritime firefighting, where decisions must be made under time pressure and uncertainty.

Maritime firefighting requires effective coordination between shipboard and shore-based actors who must maintain a common understanding of a rapidly evolving situation. This makes Shared Situational Awareness (SSA) particularly important. Endsley's three-level model describes SA through perception, comprehension and projection of a dynamic situation. The authors' previous research applied this concept to maritime firefighting and proposed an SSA-oriented Digital Solution (DS) supporting a Common Operational Picture (COP) for coordinated response (Ahmad and Merino-Laso, 2025). This raises an important educational question: how should maritime professionals be trained to interpret, integrate and act upon shared digital information during firefighting?

This question is becoming increasingly relevant as digitalization transforms maritime operations. Technologies such as Artificial Intelligence (AI), Internet of Things (IoT), Unmanned Aerial Vehicles (UAVs), and interoperable maritime information systems are changing how operational information is collected, exchanged and used. Türkistanli (2023) identifies simulator-based training as a dominant area in advanced MET research, while noting that research remains strongly concentrated on navigation-related applications. This creates an opportunity to examine advanced learning approaches in other safety-critical operations, particularly maritime firefighting.

Within OVERHEAT, previous work has progressively integrated IoT sensing, AI-supported fire detection, UAV-assisted monitoring and S-100-based maritime information into a DS for maritime fire management (Ahmad and Merino-Laso, 2025; Ahmad et al., 2025). However, richer information does not automatically improve decision-making: maritime professionals must interpret heterogeneous data, assess their relevance, maintain SA and coordinate with other actors. Manuel (2017) highlights the need for MET to prepare professionals for changing operational environments, while Mallam et al. (2019) emphasize the continuing development of technical and non-technical competencies as maritime operations become increasingly digitalized.

Simulation provides a controlled environment in which complex and hazardous situations can be reproduced while incorporating emerging digital information and collaborative workflows (Mallam et al., 2019). The present study therefore examines the educational and competency challenges arising from emerging technologies in maritime firefighting, using the OVERHEAT DS as a practical case and focusing on the knowledge, operational and technology-related competencies required for future simulator-supported MET.

# 2. RELATED WORKS

## 2.1. Evolution of Maritime Education and Training (MET)

Maritime Education and Training (MET) has evolved from practical, task-oriented training toward competency-based education combining vocational and broader academic approaches. The STCW Convention established internationally recognized competence requirements, with subsequent amendments strengthening emphasis on observable, skill- and performance-based competencies (Manuel, 2017). As maritime operations become increasingly technology-driven, MET must also prepare professionals who can adapt to changing work environments, requiring consideration of curriculum content, teaching methods, instructor capabilities, learning resources and assessment approaches (Manuel, 2017). This raises an important question for safety-critical activities such as maritime firefighting: how should mariners be prepared to integrate emerging technologies into their workflows?

## 2.2. Emerging Technologies and Digital Competencies in MET

Digitalization is progressively changing maritime operations and the competencies required from seafarers. Mallam, Nazir and Renganayagalu (2019) describe maritime operations as complex socio-technical environments requiring continuous development of technical and non-technical competencies. Technologies such as AI, IoT, UAVs and interconnected maritime information systems introduce new information that maritime professionals must interpret and incorporate into decision-making, while immersive technologies create further opportunities for MET (Mallam et al., 2019). These developments therefore affect not only how maritime tasks are performed but also how they should be taught and assessed, increasing the importance of understanding data sources, system limitations and the appropriate use of technology-generated information.

Digitalization also extends to maritime information standards. The S-100 framework provides a basis for integrating interoperable maritime information, and within OVERHEAT, S-100 services are considered alongside fire-related information and ship–shore data exchange (Ahmad and Merino-Laso, 2025). This creates an additional educational requirement: mariners need to understand how different digital information layers can be interpreted and integrated into an operational picture. S-124, as an emerging S-100-based service, provides a relevant example of this requirement, as future mariners and instructors may need opportunities to become familiar with new digital information services before wider operational adoption. Experimental and open-source environments can support such familiarization by enabling local datasets, testing and simulator-supported learning. This reinforces the need for MET to evolve alongside technological development (Manuel, 2017; Mallam et al., 2019).

## 2.3. Simulator-Based Training for Complex Maritime Operations

Simulator-based training has become an established component of MET. Its early application was particularly associated with navigation skills and emergency preparedness, including efforts to reduce accidents associated with HF (Sellberg, 2017). As simulator technology has developed in terms of fidelity, accessibility and cost-effectiveness, its application has expanded across the MET curriculum (Kim et al., 2021). Sellberg (2017) further notes that simulators are now used across different maritime activities, including bridge operations, cargo handling, engine control, crane operations and Vessel Traffic Services (VTS), while their use is incorporated into STCW requirements for simulator-based training and competency assessment.

For maritime firefighting, simulation provides a particularly valuable environment because real fire emergencies cannot be safely or repeatedly reproduced for training purposes. Controlled scenarios allow trainees to practice emergency response while developing technical and non-technical competencies such as SA, decision-making, communication and teamwork. However, the effectiveness of simulation depends not only on technological fidelity but also on how scenarios, instruction and assessment are designed. Sellberg (2017) highlights the need for greater empirical understanding of the instructional practices that support valid and reliable simulator-based education, while Kim et al. (2021) emphasize the continuum between individual skill development and more complex operational training.

The increasing digitalization of maritime operations further changes what needs to be represented in simulator-based training. Conventional simulator exercises may reproduce established operational tasks, whereas digitally enabled firefighting requires trainees to interact with new information sources and technologies, including IoT sensors, UAV thermal and optical imagery, AI-supported detection and interoperable maritime data services. The emergence of S-100-based maritime information services further extends these requirements by introducing standardized digital information layers that mariners need to understand, interpret and integrate into operational decision-making. Simulator-based environments can provide a controlled setting in which trainees become familiar with such emerging services and learn how they interact with other operational information.

Within the OVERHEAT context, this is particularly relevant to the integration of S-100 services with firefighting-related information and ship-shore data exchange, while emerging services such as S-124 may require further experimental and training environments for early familiarization. Research on maritime education has also identified limitations in conventional teaching approaches and highlighted the need for new methodologies to address learning deficiencies (Ivanišević et al., 2018; Mohović et al., 2016). Consequently, simulation can provide not only a means of practicing firefighting procedures but also an environment in which mariners learn to interpret digital information, integrate multiple data sources, maintain SA and coordinate decisions across ship-shore actors. This provides the basis for examining the OVERHEAT simulation environments as a means of identifying the knowledge, operational and technology-related competencies required for digitally supported maritime firefighting. Based on the literature, Figure 1 summarizes the link between emerging technologies, maritime firefighting competencies and simulator-supported MET.

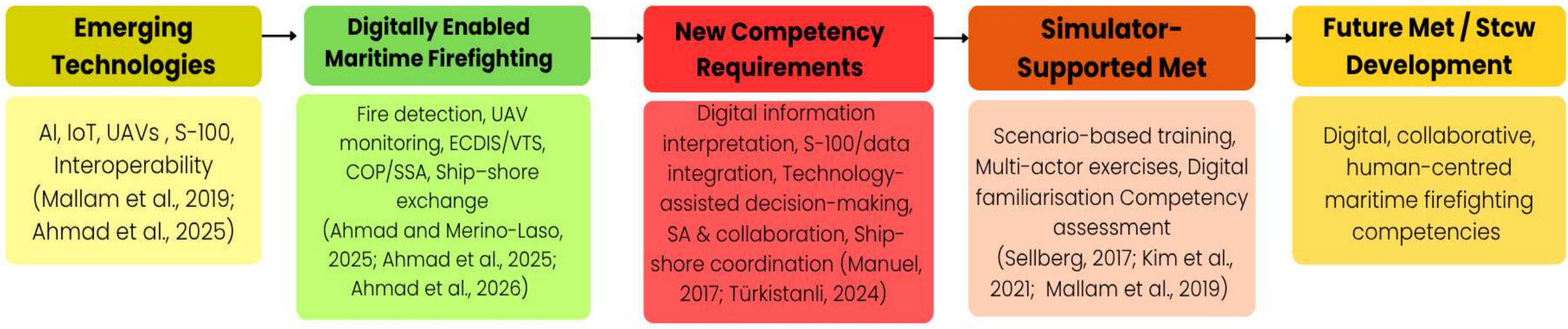

Figure 1. Conceptual Framework for Emerging Technologies and Maritime Firefighting MET

## 3. OVERHEAT'S DIGITAL SOLUTION (DS)

### 3.1. OVERHEAT DS Architecture

The OVERHEAT project develops an integrated DS, corresponding to NAV-100, to support the prevention, detection and management of containership fires through coordinated information exchange between vessel and shore-based actors. The architecture comprises onboard and shore-based components, with ECDIS serving as the vessel-side information hub and VTS, port and emergency-response functions supporting the shore-side operational picture that provides a COP for the actors involved in maritime fire management. The principal interfaces include onboard ECDIS, shore-based VTS and the Portable Pilot Unit (PPU), with functions including real-time data monitoring, drone integration, alert management and S-100 integration (Ahmad et al., 2025). The DS combines IoT sensors, UAV thermal and optical monitoring and AI-supported analytics to integrate detection, monitoring, information management and emergency coordination within a common digital environment.

### 3.2. Data Interoperability, S-100 and Shared Situational Awareness (SSA)

Interoperability is a central element of the OVERHEAT DS, which uses Service-Oriented Architecture (SOA) and the S-100 framework to integrate heterogeneous information sources. It includes IEC 61162 for simulator and navigation data, RTSP for video and JSON for metadata exchange. enabling information from different sources to be processed through interconnected maritime interfaces (Ahmad et al., 2025).

The integration of S-100 services is particularly relevant to MET. Within OVERHEAT, the S-100 layers considered include S-101 Electronic Navigational Charts, S-102 Bathymetry, S-111 Surface Currents, S-104 Water Level, S-127 Traffic Management, S-124 Navigational Warnings and S-129 Under Keel Clearance. Previous work further examined the use of S-100 to structure information from IoT sensors, UAVs, AIS and navigation systems and to support its visualization across maritime platforms (Ahmad et al., 2025).

Previous work also established the SSA-oriented basis by connecting vessel and shore information through a COP for coordinated maritime firefighting (Ahmad and Merino-Laso, 2025). Subsequent work extended the architecture through IoT, UAV, AI and S-100 integration and further addressed interoperability across ship and shore systems (Ahmad and Merino-Laso, 2025; Ahmad et al., 2025; Ahmad et al., 2026). For MET, this creates a requirement for mariners to understand information sources, interpret digital data layers, prioritize information and use a shared operational picture for collaborative decision-making. Emerging S-100 services such as S-124 may therefore also require experimental environments for familiarization and training.

### 3.3. Integration into Simulator-Supported Training

The OVERHEAT validation approach progressively combines laboratory experiments, simulator-based demonstrations and hybrid operational environments. The French use case involves a vessel in distress approaching the Port of Brest, with maritime pilots and a rescue vessel participating in the response. The configuration connects four operational terminals: the Harbour Master's VTS, the captain's onboard ECDIS, a training ECDIS connected to the ENSM simulator and a PPU for maritime pilots. The information environment includes S-100 navigation services, bathymetry, water levels, surface currents, AIS/vessel traffic information and live drone imagery.

The ENSM simulation environment includes the MARINS Full Mission Bridge Simulator (FMBS) and the open-source SOMOS simulator. MARINS comprises two bridges and a control station, but its internal architecture cannot be modified; therefore, the integration approach relies on available simulator outputs and external interfaces. This provides an important practical consideration for MET, to integrate emerging digital technologies with existing simulator infrastructures through interoperable interfaces.

The simulator-supported environment therefore combines operational simulation with the digital information environment created by NAV-100. This creates a basis for examining competencies in digital information interpretation, S-100 data integration, real-time data management, technology-assisted decision-making and ship-shore coordination (Manuel, 2017; Mallam et al., 2019; Kim et al., 2021; Türkistanli, 2023). Thus, simulator-supported validation links the technological developments of OVERHEAT with the educational requirements of future MET in maritime firefighting.

## 4. RESULTS

Initial validation activities demonstrated the ability of the OVERHEAT sensing and UAV components to provide additional information for maritime fire assessment. Controlled experiments with the IoT-enabled fire detection system demonstrated early identification of heat, smoke and gases from simulated fire conditions. UAV-based monitoring also provided thermal and optical imagery for remote observation of fire-related conditions. These results indicate the value of combining heterogeneous information sources rather than relying on a single detection mechanism. In the present study, these technological findings are interpreted from MET perspective, as creating a need to understand sensor alerts and thermal/optical information, understanding their characteristics and limitations, and integrating these inputs into the evolving fire situation.

The validation of the OVERHEAT architecture demonstrates the feasibility of integrating fire-related information with navigation and operational data through interoperable digital services. IoT, UAV, AIS and navigation information are combined through SOA and S-100-based structures, while the French configuration connects vessel, VTS, pilot and training interfaces (Ahmad et al., 2025; Ahmad et al., 2026). This supports a more synchronized operational picture across ship-shore actors and extends the SSA approach developed in the authors' previous work, in which vessel and shore information contributed to a COP for coordinated maritime firefighting (Ahmad and Merino-Laso, 2025). The educational implication derived in the present study is that the simulator-based training should require trainees to interpret shared information, prioritize heterogeneous data coordinate with other actors and maintain SSA rather than operate individual technologies in isolation.

Based on these findings, the present study identifies several initial competency requirements for future MET. Maritime professionals need to interpret heterogeneous digital information, understand how data are combined and presented through ECDIS, VTS and PPU interfaces, and apply this information during collaborative emergency response. The project also identified practical challenges affecting training integration, including simulator constraints, data-source interoperability, UAV integration and interface complexity and information overload. In addition, training activities to date have involved familiarization rather than dedicated modules for every potential user group.

| **Emerging capability** | **Educational implications derived in this study** |
|---|---|
| IoT fire sensors | Interpretation of sensor-based fire information and alerts |
| UAV thermal/optical monitoring | Interpretation and integration of remote visual information |
| AI-supported analysis | Understanding and appropriate use of technology-assisted information |
| S-100 services | Interpretation and integration of digital maritime data layers |
| ECDIS/VTS/PPU interoperability | Management of information across connected ship-shore interfaces |
| COP/SSA | SSA and collaborative decision-making |
| Multi-source data integration | Information prioritization and management of information overload |

Table 1. Emerging Technologies and Related Maritime Firefighting Training Competencies

Taken together, the results suggest that the educational challenge is not simply to train mariners to operate individual technologies, but to prepare them to integrate technology-generated information into established firefighting and emergency-response workflows. The results therefore show that the OVERHEAT findings provide the technological and operational basis, while the present study derives their implications for MET in terms of knowledge, operational and technology-related competencies.

## 5. CONCLUSION AND DISCUSSION

The findings from OVERHEAT, together with the educational analysis presented in this study, indicate that emerging technologies are changing not only the technical capabilities available for maritime firefighting but also the competencies required to use them effectively. IoT sensors, UAV-based thermal and optical monitoring, S-100 services and interconnected ECDIS/VTS interfaces create an information-rich operational environment in which maritime professionals must interpret multiple data sources, maintain SA and coordinate decisions with other actors. This is consistent with Manuel (2017), who highlights the need for MET to prepare professionals for changing operational environments, and with Mallam et al. (2019), who emphasize the continuing development of technical and non-technical competencies in increasingly digitalized maritime operations.

The findings also highlight the role of simulation in developing these competencies. Simulator-supported environments can incorporate sensor alerts, UAV imagery, S-100 information, ECDIS/VTS interfaces and shared operational information into realistic emergency scenarios. This is consistent with Kim et al. (2021) and Türkistanli (2023), who highlight the role of simulation in developing increasingly complex maritime competencies. The competencies identified in this study include digital information interpretation, heterogeneous data integration, S-100 data use, technology-assisted decision-making, SSA and ship–shore coordination. These should complement, rather than replace, established firefighting knowledge and practical skills.

The findings also have implications for future STCW-related competency development. They do not imply a direct revision of existing STCW requirements, but suggest greater emphasis on HF, digital information management, human-computer interaction (HCI), SA and collaborative decision-making as maritime operations become increasingly digitalized. Future work should translate these competencies into specific learning outcomes, simulator scenarios and assessment criteria and validate them with maritime professionals and instructors (Manuel, 2017; Mallam et al., 2019; Türkistanli, 2023).

The study is subject to the progressive development of OVERHEAT, with current activities combining laboratory validation, simulator-based work and developing demonstration environments. Limitations include simulator integration, heterogeneous data sources, UAV integration, interface complexity and information overload, while some requirements remain partially demonstrated or outside the current prototype scope.

Overall, the OVERHEAT case demonstrates that digital transformation in maritime firefighting creates corresponding requirements for MET. Future training should prepare mariners not only to operate emerging technologies but also to interpret, integrate and critically use technology-generated information within firefighting and emergency-response workflows. Simulator-supported environments offer a pathway toward more digital, collaborative and human-centred MET.

**ACKNOWLEDGEMENTS:** This work has received funding from HORIZON Europe with grant agreement No. 101076633 (OVERHEAT project) under European Union’s Horizon Europe Innovation Actions Framework Programme.